\documentclass[aps,prb,10pt,twocolumn,superscriptaddress,longbibliography,showkeys]{revtex4-2}
\usepackage[utf8]{inputenc}
\usepackage{amsmath,amssymb}
\usepackage{empheq}
\usepackage{lipsum}
\usepackage{mathtools,cuted}
\usepackage{esvect}
\usepackage{multirow}
\usepackage{xcolor}
\usepackage{soul}
\usepackage[export]{adjustbox}
\usepackage{graphicx}
\usepackage{dcolumn}
\usepackage{bm}
\usepackage[colorlinks=true,citecolor=blue]{hyperref}
\usepackage{nameref}
\usepackage{physics}
\usepackage[mathlines]{lineno}
\usepackage[capitalise]{cleveref}
\usepackage{bbm}
\usepackage{orcidlink}

\begin{document}
	
\title{Tensor network approach to momentum-resolved spectroscopy
in non-periodic super-moir\'e systems}

\author{Anouar Moustaj\,\orcidlink{0000-0002-9844-2987}}
 \affiliation{Department of Applied Physics, Aalto University, 02150 Espoo, Finland}
\author{Yitao Sun\,\orcidlink{0009-0002-9479-7147}}
 \affiliation{Department of Applied Physics, Aalto University, 02150 Espoo, Finland}
 \author{Tiago V. C. Ant\~ao\,\orcidlink{0000-0003-3622-2513}}
  \affiliation{Department of Applied Physics, Aalto University, 02150 Espoo, Finland}
\author{Jose L. Lado\,\orcidlink{0000-0002-9916-1589}}
 \affiliation{Department of Applied Physics, Aalto University, 02150 Espoo, Finland}

\date{\today}

\begin{abstract}
Computing spectral functions in large, non-periodic super-moir\'e systems remains an open problem due to the exceptionally large
system size that must be considered. Here, we establish a tensor network methodology that allows computing momentum-resolved
spectral functions of non-interacting and interacting super-moir\'e
systems at an atomistic level.
Our methodology relies on encoding an exponentially large tight-binding problem as an auxiliary
quantum many-body problem, solved with a many-body kernel polynomial tensor network algorithm combined with a quantum Fourier transform tensor network. 
We demonstrate the method for one and two-dimensional
super-moir\'e systems, 
including super-moir\'e with
non-uniform strain,
interactions treated at the mean-field level,
and quasicrystalline
super-moir\'e patterns.
Furthermore, we demonstrate that our
methodology
allows us to compute
momentum-resolved spectral functions restricted to selected regions of a super-moir\'e,
enabling direct imaging of 
position-dependent electronic structure
and minigaps
in super-moir\'e systems with non-uniform strain.
Our results establish a powerful methodology to compute momentum-resolved spectral functions in exceptionally large
super-moir\'e systems, providing
a tool to directly model quantum twisting microscope experiments in twisted van der Waals heterostructures.
\end{abstract}
	
\maketitle

\section{Introduction}\label{sec:Intro}
The stacking of van der Waals materials\cite{Geim2013VanHeterostructures,Andrei2021}, giving rise to moir\'e physics, offers a versatile framework for exploring and engineering a broad spectrum of correlated phases \cite{Ahn2018DiracQuasicrystal,Cao2021NematicityGraphene,Li2021QuantumBands,Ramires2021EmulatingGraphene,Ruan2021EvidenceMicroscopy,Kezilebieke2022Moire-EnabledSuperconductivity,Uri2023SuperconductivityQuasicrystal,Park2025EvidenceSuperlattice, Park2021,Lu2019,Zhang2022,Yankowitz2019,Klein2023,Zeng2023,Cai2023,Serlin2020,Zeng2023,Lu2024,Cao2018b,Zhao2023,Kim2023,Vao2021,Zhao2023}. 
However, their theoretical modeling remains an open challenge. 
While continuum models
allow modeling uniform moir\'e systems at relatively low cost \cite{PhysRevLett.99.256802,PhysRevB.88.121408,PhysRevB.99.205134,Bistritzer2011MoireGraphene,PhysRevLett.122.106405,PhysRevX.8.031087},
defects such as impurities or domain walls
represent a challenge for them.
Real-space models provide an alternative to
study defects \cite{PhysRevB.99.245118,PhysRevMaterials.3.084003,PhysRevB.108.125141,Ramzan2023} and non-uniformity in moir\'e materials \cite{Manesco2021,PhysRevLett.128.176406}.
However the required systems sizes for real-space methods
very quickly
reach hundreds of thousands of sites,
becoming challenging for conventional methodologies. Beyond moir\'e systems, stacking multiple layers can give rise to moir\'e-of-moir\'e, or super-moir\'e physics, 
leading to new emergent exotic phenomena \cite{Ahn2018DiracQuasicrystal,Uri2023SuperconductivityQuasicrystal}.
Super-moir\'e materials characteristic length scales become significantly larger
than moir\'e systems, quickly reaching millions or even billion sites \cite{Chen2019EvidenceSuperlattice,Zhu2020ModelingHeterostructures}.

The development of quantum twisting microscope (QTM) \cite{Inbar2023TheMicroscope,Xiao2023ProbingInterferometry,Peri2024ProbingMicroscope,Xiao2024TheoryMicroscope,Xiao2024TheoryMicroscope,Birkbeck2025QuantumGraphene} have enabled local, momentum-resolved measurements of spectral functions in two-dimensional van der Waals materials. Unlike angle-resolved photoemission spectroscopy (ARPES) \cite{Zhang2022Angle-resolvedSpectroscopy,Cattelan2018AMaterials}, which averages over large illuminated regions, the QTM accesses momentum space locally via coherent tunneling across a nanoscale, twist-controlled interface. These novel experimental capabilities motivate the development of numerical methods that can bypass conventional memory bottlenecks and faithfully reproduce momentum-resolved observables in the extremely large-scale, inhomogeneous moir\'e and super-moir\'e systems now accessible experimentally.

While solving large-scale tight-binding models is a major computational challenge \cite{Joo2020}, a recent strategy leveraging quantum many-body 
solvers has enabled solving exponentially large single-particle problems \cite{Fumega2025CorrelatedAlgorithm,Sun2025,Antao2025TensorMosaics}. This technique circumvents the need to store large matrices by
leveraging tensor networks,
a many-body methodology that enables performing algebraic operations
in exponentially large spaces \cite{PhysRevLett.69.2863,RevModPhys.93.045003,Bauls2023,Ors2014,Schollwck2011,Stoudenmire2012,Fishman2022TheCalculations,Fishman2022CodebaseITensor,PhysRevX.10.041038,Huggins2019,Ors2019,Chan2016},
combined with tensor cross interpolation \cite{Oseledets2010,Oseledets2011,Ritter2024QuanticsFunctions,PhysRevB.110.035124,NunezFernandez2025LearningLibraries,QuanticsTCI.jl,TensorCrossInterpolation.jl}
to build the required tensor networks. 
The strategy of using tensor networks to represent classical functions
has enabled speedups of several orders of magnitude
in a variety of computational problems \cite{2026arXiv260103035W,PhysRevLett.132.056501,PhysRevX.13.021015, PhysRevB.107.245135,PhysRevX.12.041018,10.21468/SciPostPhys.18.1.007,PhysRevB.110.035124,Jeannin2025},
including computational chemistry~\cite{Jolly2025}, and dynamics problems \cite{Peddinti2024,Gourianov2025,niedermeierGP,boucomasGP,chenGP,2025arXiv250701149C}. 
This enabled the study of quasicrystalline mosaics of topological Chern states \cite{Antao2025TensorMosaics},
and solving interacting super-moir\'e systems self-consistently in system sizes above one billion sites \cite{Sun2025}. 
However, current tensor-network–based tight-binding methods are inherently formulated in real space and do not yet permit direct computation of momentum-resolved spectral functions, thereby limiting their applicability to QTM and ARPES measurements in  super-moir\'e structures.

In this work, we introduce a methodology that enables 
computing spectral functions in momentum space
in super-moir\'e materials with tensor networks,
providing direct access to band structures and the prediction of observables probed by QTM or ARPES measurements. Our algorithm
exploits a tensor-network representation of
the quantum Fourier transform \cite{Shor1997Polynomial-TimeComputer,Kitaev1995QuantumProblem,Chen2023QuantumEntanglement,Chen2025DirectOperator} (QFT),
enabling reconstructing the momentum-resolved spectral function. 
We demonstrate our approach
in super-moir\'e systems in one and two dimensions (1D and 2D),
featuring interactions, inhomogeneous strain,
and quasicrystalline patterns.
Furthermore, we show that our methodology enables to compute
the local momentum-resolved electronic structure,
allowing to image local changes to the bandstructure
induced by non-uniformity.
Our manuscript is organized as follows. In \cref{Sec: Methods}, we
introduce the tensor-network formulation of tight-binding models,
and the strategy based on a quantum Fourier transform
for momentum resolved spectra functions, both in the presence
and absence of interactions.
In \cref{Sec: Results}, we demonstrate our methodology for two non-periodic
cases in 1D and 2D. 
Finally, in \cref{Sec: Conclusion}, we summarize our results.

\section{Methods}\label{Sec: Methods}\label{sec: Tensor TB models}

\begin{figure}[!t]
    \centering
    \includegraphics[width=\columnwidth]{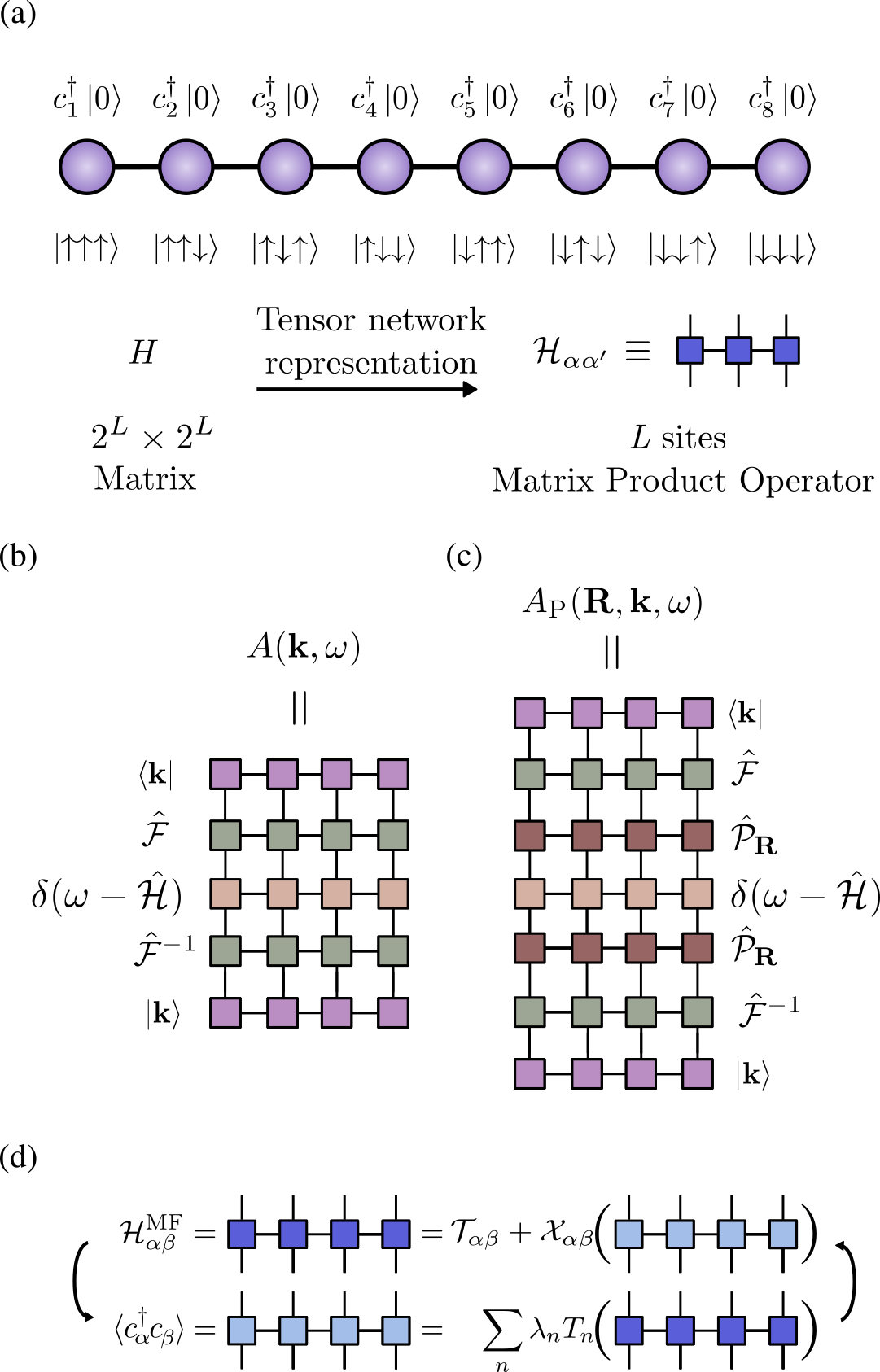}
    \caption{(a) Schematic of the mapping between
    a single particle problem with $N=2^L=8$ sites,
    and a many-body pseudo-spin chain of length $L=3$. While a sparse $2^L\times2^L$ matrix represents the Hamiltonian in the real-space representation, it is represented by an $L$-site MPO in the tensor-network representation. (b) Tensor network algorithm to compute the momentum-resolved spectral function. The purple MPSs at the top and bottom of this network represent momentum basis states $\ket{\mathbf{k}}$. The green MPOs represent the Quantum Fourier Transform $\hat{\mathcal{F}}$ and its inverse $\hat{\mathcal{F}}^{-1}$, which are acting on the
    operator $\delta (\omega-\hat{\mathcal{H}} )$. (c) The same tensor network as in (b), but augmented with projection MPOs $\hat{\mathcal{P}}_\mathbf{R}$ to resolve the spectral functions locally. (d) The tensor-network SCMF loop. Here, $\mathcal{T_{\alpha\beta}}$ represents the non-interacting part of the tensorized Hamiltonian and $\chi_{\alpha\beta}$ represents the one-body operator in tensorized form that results from iteratively computing $\langle c_\alpha^\dagger c_\beta\rangle$ in the Hubbard term in \cref{Eq: Hubbard term}. }
    \label{fig: Methods Summary Fig}
\end{figure}

We consider a generic tight-binding Hamiltonian of the form
\begin{equation}\label{Eq: Gen Hamiltonian}
\begin{split}
    \hat{H}_0 &= 
             \sum_{\alpha, \beta,\sigma}h_{\alpha \beta}c^\dagger_{\alpha\sigma}c_{\beta\sigma} , 
\end{split}
\end{equation}
where $h_{\alpha \beta}$ are the matrix elements of the Hamiltonian in the single-particle basis
and $c^\dagger_{\alpha\sigma},c_{\alpha\sigma}$
are the creation/annihilation operators for site $\alpha$ and spin $\sigma$.
The instrumental step in our methodology is to represent all the required operations
in an exponentially large tight-binding model using tensor network algorithms,
as we elaborate below.

\subsection{Tensor-network representation tight-binding Hamiltonians}

To convert the real-space tight-binding Hamiltonian
into a tensor-network Hamiltonian, we will perform a pseudo-spin encoding
of the lattice indices via $\alpha=(s_1,s_2,...,s_L)$, with $s_\alpha = \uparrow,\downarrow$ 
In this form, the original single particle Hamiltonian for $N=2^L$ sites $H_{\alpha\beta}$
becomes a many-body Hamiltonian
of $L$ pseudo-spins in the basis $\ket{\alpha} = \ket{s_1,s_2,...,s_L}$,
$\mathcal{H}_{\alpha\alpha'} = \langle \alpha | \hat{\mathcal{H}} |\alpha'\rangle =
\mathcal{H}_{(s_1,s_2,...,s_L),(s'_1,s'_2,...,s'_L)}$.
In this pseudo-spin basis, the Hamiltonian can
be written as a matrix product operator (MPO) as $\mathcal{H}_{(s_1,s_2,...,s_L),(s'_1,s'_2,...,s'_L)} = {\Gamma}^{s_1,s_1'}_1{\Gamma}^{s_2,s_2'}_2\cdots {\Gamma}^{s_L,s_L'}_L$.
The tensors $\Gamma_r$ are four-indexed tensors, where two indexes are virtual
and contracted with adjacent tensors, while the remaining two are physical, corresponding to the local two-dimensional Hilbert space $s_r,s'_r$. 
The virtual indexes have dimension $\chi$, known as
the bond dimension, which controls the complexity of the local tensor,
which in the language of many-body quantum systems is a measure of the entanglement. This process is depicted in \cref{fig: Methods Summary Fig}(a), where the real-space electronic system is a chain of length $N=8$ and the tensorized Hamiltonian becomes a many-body pseudo-spin chain of length $L=\log_2(N)=3$. Each of the eight basis elements of the single-particle Hamiltonian becomes a basis element of a smaller system of spins. 
This formulation is analogous to the quantics framework \cite{Ritter2024QuanticsFunctions}, which encodes the grid on which a function is numerically approximated via a binary expansion of the grid points, and subsequently encodes the function values as multi-index tensors. This structure enables the application of tensor cross interpolation (TCI) techniques for compact storage and efficient tensor contractions\cite{itensor}, thereby facilitating the representation of operations on such functions \cite{NunezFernandez2025LearningLibraries}. Moreover, it allows us to use the broader class of Quantics tensor cross interpolation (QTCI) algorithms developed within this formalism
to represent highly featured modulations on the lattice \cite{Ritter2024QuanticsFunctions}. 

For clarity, we illustrate the construction for a one-dimensional, non-interacting, spinless tight-binding Hamiltonian $\hat{H}$. In real space, $\hat{H}$ can be decomposed into a kinetic matrix $\hat{K}$, which is a shift matrix with non-zero elements $\bra{i}\hat{K}\ket{i+1} = 1$, and a diagonal matrix $\hat{T}(1) = \text{diag}(t^{(1)}_1, t^{(1)}_2, \cdots)$ containing the (potentially spatially varying) hopping amplitudes. Together, they define the upper triangular part of the Hamiltonian as $\hat{T}\hat{K}$. 
Higher-order neighbor hoppings are naturally constructed by successive powers of $\hat{K}$, each weighted by a corresponding diagonal matrix $\hat{T}(n) =\text{diag}(t^{(n)}_1, t^{(n)}_2, \cdots)$, yielding the general form $\hat{H} = \sum_n \hat{T}(n)\hat{K}^n + \text{h.c.}$.  
To express this Hamiltonian in tensor form, we construct an equivalent many-body kinetic operator that reproduces the action of $\hat{K}$ on the pseudo-spin basis states. This is achieved by using the operator $\hat{\mathcal{K}} = \sum_{r=1}^L \sigma_r^+ \bigotimes_{m > r} \sigma_m^-$, where $\sigma_r^{\pm} = (\sigma_r^x \pm i \sigma_r^y)/2$ are Pauli ladder operators acting at site $r$. The hopping amplitudes are then encoded in diagonal MPOs, constructed from matrix product states (MPS) generated using the QTCI algorithm applied to an arbitrary function $t(x_i)$.
The full tensorized Hamiltonian therefore takes the general form
\begin{equation}
\hat{\mathcal{H}} = \sum_n \hat{\mathcal{T}}(n)\hat{\mathcal{K}}^n + \text{h.c.},
\end{equation}
where each term describes the $n^\text{th}$-neighbor hopping, with $\hat{\mathcal{K}}^n$ encoding the shift and $\hat{\mathcal{T}}(n)$ the spatially varying amplitudes. The $n=0$ term represents on-site modulated potentials that can likewise be implemented as diagonal MPOs constructed via QTCI (see \cite{Sun2025,Antao2025TensorMosaics,Kazeev2012,Kazeev2013} for further details).

Once the Hamiltonian is obtained, we can compute spectral quantities using the kernel polynomial method (KPM) \cite{Weie2006TheMethod}. For instance, the spectral function is calculated from the Dirac-delta operator, which takes the form
\begin{equation}\label{Eq: delta expansion}
    \delta(\omega - \hat{\mathcal{H}}) = \frac{1}{\pi\sqrt{1-\omega^2}}\left[\mathbbm{1} + 2\sum_{n=1}^\infty\hat{\mu}_nT_n(\omega)\right],
\end{equation}
where $T_n(x)$ is a Chebyshev polynomial satisfying the recurrence relation $T_0(x) = 1$, $T_1(x) = x$, and $T_n(x) = 2 T_{n-1}(x) - T_n(x)$ for $n>1$, with $\omega$ and $\hat{\mathcal{H}}$ being rescaled frequencies and Hamiltonians, as the KPM requires functions taking values in the domain $D = (-1,1)$.
The Chebyshev moments for this operator are simply given by $\mu_n = \int_{-1}^{1}d\omega\delta(\omega-\hat{\mathcal{H}})T_n(\omega) = T_n(\hat{\mathcal{H}})$,
which can be directly computed using the Chebyshev recurrence relations
with matrix product operators. In practice, only $N_\mu$ terms are retained, as the KPM, when combined with the Jackson kernel \cite{Jackson1912}, leads to an effective energy smearing of order $1/N_\mu$ that controls the spectral broadening \cite{Weie2006TheMethod}. 
From this, one can obtain, for instance, the local density of states (LDOS) $\rho(\mathbf{r},\omega)=\bra{\mathbf{r}}\delta(\omega-\hat{\mathcal{H}})\ket{\mathbf{r}}$, or the momentum-resolved spectral function $A(\mathbf{k}, \omega)$, using a KPM tensor-network algorithm \cite{Weie2006TheMethod}. The main object of this work is the
momentum-resolved spectral function, defined as 
\begin{equation}\label{Eq: Mom-Res Spc Func}
    A(\mathbf{k},\omega) = \bra{\mathbf{k}}\hat{\mathcal{F}}\delta\left(\omega-\hat{\mathcal{H}}\right)\hat{\mathcal{F}}^{-1}\ket{\mathbf{k}},
\end{equation}
where $\hat{\mathcal{F}}$ is the Fourier transform that in the pseudo-spin representation becomes a Quantum Fourier Transform (QFT), as
we elaborate on in the following subsection. The tensor network giving rise to $A(\mathbf{k}, \omega)$ is depicted in \cref{fig: Methods Summary Fig}(b).

\subsection{The Quantum Fourier Transform}

The momentum-resolved spectral function $A(\mathbf{k}, \omega)$ is obtained by using the QFT operator \cite{Chen2023QuantumEntanglement} in the many-body pseudo-spin representation of the full spectral function. 
The QFT notably underpins Shor’s factorization algorithm \cite{Shor1997Polynomial-TimeComputer} and is also used in the quantum phase estimation problem in quantum circuits \cite{Kitaev1995QuantumProblem}. 
A central key point of our algorithm is that the QFT can be efficiently
represented as a tensor network with very low bond dimension \cite{Woolfe2017,Roberts2014,Chen2023QuantumEntanglement,Chen2025DirectOperator}.
It is worth to first elaborate on
how we can leverage a quantum algorithm for a classical computation.
For the sake of concreteness, we focus the discussion
in a one-dimensional system,
noting that the procedure can be readily extended to two dimensions.
For single particle basis states in real space $\ket{\alpha}$, the QFT is equivalent in definition to the discrete Fourier transform, which acts as 
$\hat{F}:\ket{\alpha} \mapsto \ket{k}$ with
\begin{equation*}
    \ket{k} = \frac{1}{\sqrt{N}}\sum_{\alpha=1}^N\omega_N^{k\alpha}\ket{\alpha}, \ \ \omega_N^{k\alpha}=e^{2\pi ik\alpha/N}.
\end{equation*}
Note that this transform can be applied to any discrete set of basis states, irrespective of whether the underlying system is periodic. Incommensurability enters solely through the Hamiltonian matrix elements $H_{x,x'}$, for instance via quasiperiodic onsite modulations, while the basis itself remains unaffected. In the absence of periodicity, momentum ceases to be a conserved quantum number, and consequently, spectral functions in the Fourier basis are generally non-diagonal. The QFT therefore provides a reciprocal-space representation rather than a set of translation eigenstates. As a unitary transformation, it preserves all information, so the full real-space structure, including any incommensurate features, remains encoded in the transformed representation.

The binary representation of the single particle basis states can be written as $\ket{\alpha}=\ket{s_1 s_2\cdots s_L}$, where $s_r=\downarrow,\uparrow \equiv 0,1$ can be understood as the binary representation of $\alpha=\sum_r s_r 2^{L-r}$. It then follows that one can write
$\omega_N^{k\alpha}=\prod_re^{2\pi iks_r2^{-r}}$, or 
\begin{align*}
   \hat{F} \ket{k} &=\frac{1}{\sqrt{N}}\bigotimes_{r=1}^L\sum_{s_r=0}^1e^{2\pi iks_r2^{-r}}\ket{s_r} \\
            &= \bigotimes_{r=1}^L\frac{1}{\sqrt{2}}\left[\ket{0} + e^{2\pi ik2^{-r}}\ket{1}\right]
\end{align*}
By expanding the basis vector $\ket{k}=\ket{\bar s_1\bar s_2\cdots\bar s_L}$, it can be shown that the full representation of the Fourier transform action on the many-body pseudo-spin basis (represented here as qubits) is given by
\begin{equation}\label{Eq: Qubit QFT}
\begin{split}
        &\hat{\mathcal{F}}\ket{\bar s_1\bar s_2\cdots\bar s_L} =\frac{1}{\sqrt{2^L}}\begin{multlined}[t]
             \left[\ket{0} + e^{2\pi i[0.\bar s_L]}\ket{1}\right]\otimes\\
             \left[\ket{0} + e^{2\pi i[0.\bar s_{L-1}\bar s_L]}\ket{1}\right]\otimes \\
             \cdots\otimes\left[\ket{0} + e^{2\pi i[0.\bar s_1\bar s_2\cdots\bar s_L]}\ket{1}\right],
        \end{multlined}
\end{split}
\end{equation}
where we used fractional binary notation $[0.\bar s_m\bar s_{m+1}\cdots\bar s_L]=\sum_{r=m}^L\bar s_r2^{-r}$, with $m\in\{1,\cdots,L\}$. This unitary operation can be implemented as a quantum circuit by the application of a Hadamard gate $\hat{H}$ and a sequential application of the controlled dyadic rational phase gate $C\hat{R}_\kappa$. These gates are defined as 
\begin{equation*}
    \hat{H} = \frac{1}{\sqrt{2}}\begin{pmatrix}
        1 & 1 \\
        1 & -1
    \end{pmatrix}, \ \ \hat{R}_\kappa = \begin{pmatrix}
        1 & 0 \\
        0 & e^{2\pi i2^{-\kappa}}
    \end{pmatrix}.
\end{equation*}
The first step of the circuit implementation is
\begin{align*}
    &\text{(1)} \ \ \hat{H}_1\ket{\underline{\bar s}}= \frac{1}{2}\left(\ket{0} + e^{2\pi i [0.\bar s_1]}\ket{1}\right)\ket{\bar s_2\cdots}, \\
    &\text{(2)}  \ \ \hat{R}_2\hat{H}_1\ket{\underline{\bar s}} =  \frac{1}{2}\left(\ket{0} + e^{2\pi i [0.\bar s_1\bar s_2]}\ket{1}\right)\ket{\bar s_2\cdots}, \\
    & \ \ \vdots \\
    &\text{($L$)} \ \ \hat{R}_L\cdots\hat{R}_1\hat{H}_1\ket{\underline{\bar s}} = \frac{1}{2}\left(\ket{0} + e^{2\pi i [0.\bar s_1\cdots\bar s_L]}\ket{1}\right)\ket{\bar s_2\cdots},
\end{align*}
where we used the shorthand notation $\ket{\underline{\bar{s}}}=\ket{\bar s_1 \bar s_2 \cdots \bar s_L}$. This procedure costs $L$ operations and needs to be repeated for the remaining $L-1$ qubits, followed by $L/2$ swap operations to yield \cref{Eq: Qubit QFT}. Thus, the QFT is a unitary transformation equivalent to the discrete Fourier transform but with a reduced computational complexity of $\mathcal{O}(L^2)$, where $L=\log_2N$, which is much cheaper than the $\mathcal{O}(N\log N)$ operations needed to perform the fast Fourier transform. In practice, the algorithm is not implemented via the explicit circuit description shown here, but rather through optimized tensor-network interpolative schemes \cite{Chen2025DirectOperator}, implemented with QTCI \cite{NunezFernandez2025LearningLibraries}.

\subsection{Tensor-Network Self-consistent mean field calculation} \label{Sec: SCMF Ham}
Interaction effects can be treated
at the mean-field level
by performing a self-consistent density decoupling using tensor network
techniques \cite{Fumega2025CorrelatedAlgorithm,Sun2025}.
For the sake of concreteness, we take a local Hubbard interaction \cite{Arovas2022TheModel} of the form
\begin{equation}\label{Eq: Hubbard term}
    \hat{H}_{\text{int}} =\sum_{\alpha}U_\alpha\left(\hat{n}_{\alpha\uparrow}-\frac{1}{2}\right)\left(\hat{n}_{\alpha\downarrow}-\frac{1}{2}\right),
\end{equation}
where $\hat{n}_{\alpha,\sigma} = c^\dagger_{\alpha,\sigma} c_{\alpha,\sigma}$ and $U_\alpha$ is a local Hubbard interaction, which could also depend on the position $\mathbf{r}_\alpha$.
To perform the self-consistent mean-field (SCMF) decoupling, the Hamiltonian is expanded as
\begin{align*}
    \hat{H}_{\text{int},\alpha}/U_\alpha\approx &\left(\hat{n}_{\alpha\uparrow}-\frac{1}{2}\right)\left(\langle \hat{n}_{\alpha\downarrow}\rangle-\frac{1}{2}\right) + \\ &\left(\langle \hat{n}_{\alpha\uparrow}\rangle-\frac{1}{2}\right)\left(\hat{n}_{\alpha\downarrow}-\frac{1}{2}\right)
\end{align*}
where the expectation value of the density operator is taken with respect to the the many-body ground state $\ket{\Omega} = \prod_{\epsilon_\mu<\epsilon_F}\Psi^\dagger_\mu\ket{0}$,
where $\hat{\mathcal{H}}_\text{MF} | \Psi_\mu \rangle = \epsilon_\mu |\Psi_\mu \rangle$
and the eigenstates of the single-particle mean-field
Hamiltonian, and $\epsilon_F$ is the Fermi energy
and $|0\rangle$ the empty state. This can be written as 
\begin{equation}\label{Eq: mean-field density}
\begin{split}
    \langle\hat{n}_{\alpha\sigma}\rangle &= \int_{-\infty}^{\epsilon_F} d \omega \bra{0}c_{\alpha\sigma}\delta\left(\omega - \hat{\mathcal{H}}_{\text{MF}}\right)c^\dagger_{\alpha\sigma}\ket{0}, 
\end{split}
\end{equation}
or more compactly as $\langle\hat{n}_{\alpha\sigma}\rangle = \bra{\alpha,\sigma}\hat{\Xi}\left(\hat{\mathcal{H}}_{\text{MF}}\right)\ket{\alpha,\sigma}$. Here, $\hat{\mathcal{H}}_\text{MF}$ is the mean-field Hamiltonian and $\epsilon_F$ the Fermi energy. 
The operator $\hat{\Xi}$ is expanded in terms of Chebyshev polynomials, $\hat{\Xi}\left(\hat{\mathcal{H}}_{\text{MF}}\right) = \sum_{n=0}^\infty\lambda_nT_n(\hat{\mathcal{H}}_\text{MF})$,
whose moments are given by 
$
    \lambda_0 = \int_{-1}^{\Tilde{\epsilon}_{F}}\dd \omega\frac{1}{\pi\sqrt{1-\omega^2}} $ and $
    \lambda_{n\geq 1} = \int_{-1}^{\Tilde{\epsilon}_{F}}\dd\omega\frac{2T_n(\omega)}{\pi\sqrt{1-\omega^2}},
$
where $\Tilde{\epsilon}_F$ is a rescaled Fermi energy.
The mean-field Hamiltonian $\hat{\mathcal{H}}_\text{MF}$ is computed as a tensor-network self-consistent equation\cite{itensor,Sun2025}, 
starting from an initial guess for $\hat{n}_{\alpha,\sigma}$ and iterating \cref{Eq: mean-field density} until convergence is reached.
Throughout this process, the densities are taken to be diagonal MPOs, evaluated using QTCI. The local potential $U$ can also be modulated and expressed as an MPO by using the QTCI algorithm (see \cite{Sun2025} for more detail). This could be useful for modeling a spatially dependent local interaction, which is achievable through Coulomb engineering \cite{Raja2017}. The KPM SCMF process is depicted in \cref{fig: Methods Summary Fig}(c).

\section{Momentum-resolved spectral function in super-moir\'e systems}\label{Sec: Results}

\begin{figure*}[!t]
    \centering
    \includegraphics[]{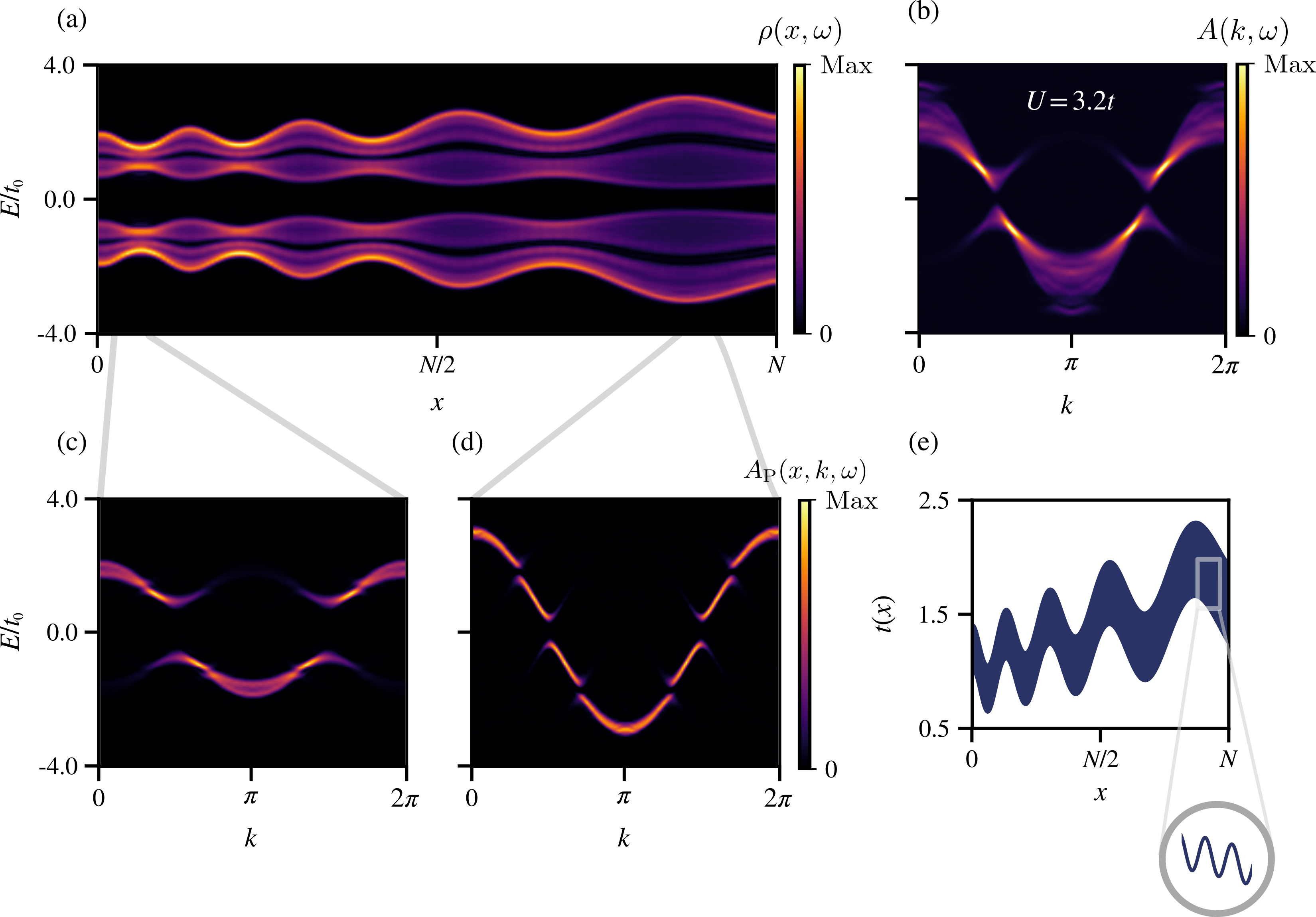}
    \caption{Simulation results in 1D. The system size is $N=2^{24}$, and the projected regions have sizes $N_X=N/16$. (a) Local density of states $\rho(x_i,\omega)$ and (b) the total momentum-space spectral function $A(k,\omega)$. (c,d) projected spectral function $A_P(k,\omega)$ for a 1D chain featuring a hopping modulation with a linearly increasing frequency and amplitude. (e) The hopping modulation shows that the amplitude, the average, the atomic-scale and the moir\'e wavelengths increase. In (b) and (c), the splitting of the band due to the incommensurate modulation is not as obvious as in (d).}
    \label{fig: projected spc func intqc1d}
\end{figure*}

In practical super-moir\'e systems, we are interested in knowing the momentum-resolved spectral function in a local region of space. This capability is essential in super-moir\'e systems where the moir\'e pattern varies slowly across the sample, resulting in local variations of the band structure. Such a spatially dependent dispersion can directly be probed by the QTM \cite{Inbar2023TheMicroscope,Birkbeck2025QuantumGraphene}, which measures the local spectral function with momentum resolution. For instance, in recently realized strain-programmable moir\'e systems \cite{Kapfer2023ProgrammingMaterials}, where both strain and moir\'e wavelength vary continuously across the sample, the QTM can reveal position-dependent band features. 
An important reformulation of \cref{Eq: Mom-Res Spc Func} is the incorporation of real-space projectors, which enable to compute the momentum-resolved spectral
function in a selected region.
Our approach captures these variations by computing the locally projected spectral function, thereby providing theoretical access to the spatial structure of momentum-resolved observables.
For this purpose, we construct MPOs $\hat{\mathcal{P}}_\mathbf{R}$ representing projectors onto the closest $N_\mathbf{R}$ sites around 
a location $\mathbf{R}$.
The projected spectral function is then defined as

\begin{equation*}\label{Eq: Mom-Res Spc Func Proj}
    A_{\text{P}}(\mathbf{R},\mathbf{k},\omega) = \bra{\mathbf{k}}\hat{\mathcal{F}}\hat{\mathcal{P}}_\mathbf{R}\delta\left(\omega-\hat{\mathcal{H}}_\text{MF}\right)\hat{\mathcal{P}}_\mathbf{R}\hat{\mathcal{F}}^{-1}\ket{\mathbf{k}}.
\end{equation*}

To exemplify the tools developed in this work, we present results in both 1D and 2D where aperiodicity and incommensurability play a significant role. 

\subsection{One-dimensional super-moir\'e and inhomogeneous strain}
\label{Sec: One-dimensional incommensurate system under inhomogeneous strain}

We consider a super-moir\'e system with non-uniform strain, leading to a modulated nearest-neighbor hopping amplitude that varies linearly in a 1D chain. The hopping amplitude is modulated by two incommensurate wavelengths and is given by
\begin{equation}\label{Eq: lin increasing modulation}
\begin{split}
    t_{i,i+1} &= t(x_i)\left[1 + \frac{1}{5}\sum_{l=1}^2\cos(\Tilde{k}_l(\Tilde{x}_i)\Tilde{x}_i)\right]
\end{split}
\end{equation}
where $\Tilde{x}_i=(x_i + x_{i+1})/2$ and $\Tilde{k}_l(\Tilde{x}) = 2\pi/\lambda_l(\Tilde{x})$. 
For the sake of concreteness, we take the interatomic distance $a=1$.
The atomic-scale wavelength is kept constant $\lambda_1(x) = \sqrt{5}/2$, while the moir\'e one grows linearly, $\lambda_2(x) = \sqrt{3}N/15(1+x/N)$. Additionally, the overall amplitude is also modulated with a linear function $t(x) = t_0(1 + 3x/4N)$, which allows us to expose the differences between the region at the start from the region at the end clearly.

\begin{figure*}[!hbt]
    \centering
    \includegraphics[width=\textwidth]{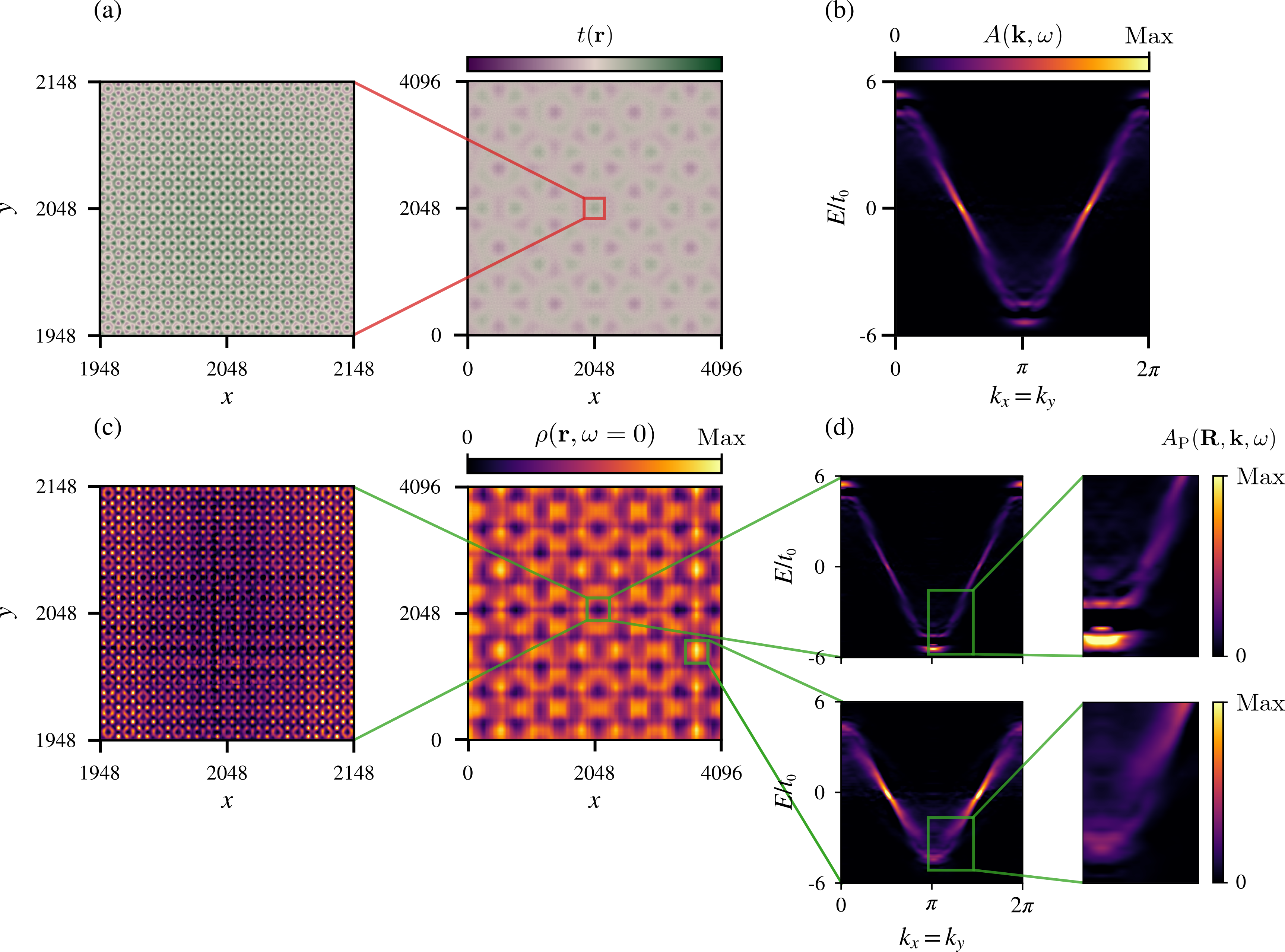}
    \caption{Simulation results in 2D. The system size is $N=2^{24}$, with $N_x=N_y=2^{12}$, and the projected regions have linear sizes $N_{\mathbf{R}}\approx N_x/16=N_y/16$. (a) The hopping function in real space and a zoom into the central region, where the two moir\'e modulations at the different scales are clearly visible. (b) The total momentum-resolved spectral function $A(\mathbf{k},\omega)$ along the line $k_x=k_y$. (c) The LDOS $\rho(\mathbf{r}.\omega=0)$ and a zoom into the same region as (a), where the spatial pattern closely follows that of the hopping modulation. (d) The projected momentum-resolved spectral function $A_\text{P}(\mathbf{R}.\mathbf{k},\omega)$ in the two regions shown in (c), with insets showing pronounced differences between them. In particular, moir\'e minigaps are more clearly resolved in the top inset.}
    \label{fig: 8QC hopping & Bands}
\end{figure*}

We show in \cref{fig: projected spc func intqc1d}(a), the local density of states $\rho(x, \omega) = \bra{x} \delta(\omega - \hat{\mathcal{H}}) \ket{x}$ along the chain, where the spectral weight clearly follows the large scale modulation pattern, reflecting the increasing frequency and amplitude. In \cref{fig: projected spc func intqc1d}(b) the corresponding total spectral function $A(k, \omega)$ is shown,
featuring a smearing out from the mixing of spectral weight from the different regions of the system. In order to resolve those features, the projected spectral functions $A_{\text{P}}(X,k,\omega)$ for two spatial regions of sizes $N_X\approx N/16$ (indicated by gray connectors), are presented in \cref{fig: projected spc func intqc1d}(c–d). Distinct variations in $A_{\text{P}}(X,k,\omega)$ between these regions are visible. Notably, the mini gaps are more pronounced in \cref{fig: projected spc func intqc1d}(d), corresponding to a region where the hopping modulation wavelength is two times larger and the amplitude is $75\%$ larger than that of the region shown in \cref{fig: projected spc func intqc1d}(c). These spatial variations are illustrated in \cref{fig: projected spc func intqc1d}(e), where the modulation function $t(x)$ is plotted. The interaction strength is set to $U = 3.2t$ and the system size to $N = 2^{24}$, far beyond the capabilities of dense matrix solvers.

\subsection{Momentum-resolved spectral function in a two-dimensional quasicrystal}

We now consider a non-interacting 2D square-lattice model with an 8-fold rotationally symmetric quasicrystalline hopping modulation potential.
Such Hamiltonians naturally arise in platforms where quasiperiodic potentials are engineered through controlled interference of multiple periodic structures. In ultracold atoms, for example, eightfold rotationally symmetric optical potentials are created by superimposing four mutually detuned standing-wave lasers at $45^\circ$ angles \cite{Viebahn2019Matter-WaveLattice,Yu2024,PhysRevLett.125.200604}. Related experiments demonstrate that such optical configurations yield dense, scale-invariant diffraction patterns and effectively realize higher-dimensional tight-binding models via the cut-and-project mechanism \cite{PhysRevLett.125.200604,Gottlob2023HubbardPotentials}.  
Here, to demonstrate the capabilities of our method, we
include quasicrystalline
incommensurate modulations at two widely different
length scales. 
To this end, we take a system where the hopping amplitude is given by
\begin{equation}\label{Eq: U potential 8fold}
    t(\mathbf{r}) = t_0\left[1 + \sum_{n=1}^4\left(\Delta_\alpha \cos\left(\alpha \mathbf{k}_n\cdot \Tilde{\mathbf{r}}\right) + \Delta_\beta \cos\left(\beta \mathbf{k}_n\cdot \Tilde{\mathbf{r}}\right)\right)\right],
\end{equation}
where $\Delta_\alpha$ controls the modulation strength at the atomic scale,
$\Delta_\beta$ controls the modulation strength at a much bigger length scale,
$\mathbf{k_n}=R^n(\pi/4)\left[2\pi, 0\right]^T$, with $R(\pi/4)$ being the 2D rotation matrix with an angle of $\pi/4$ radians. We also introduced a short-hand notation for shifted coordinates $\Tilde{\mathbf{r}}_i\equiv \left[x_i-N_x/2, y_i - N_y/2\right]^T$ such that the rotationally invariant point lies at the center of the lattice. 
We take $\Delta_\alpha = 0.25$, $\Delta_\beta = 0.1$ and moir\'e scales $\alpha = 1/10\sqrt{2}$ and $\beta = 16/\sqrt{3}N_x $. The hopping amplitude is always evaluated at the midpoint between nearest-neighbors, $\mathbf{r}_{i,i+1} = (\mathbf{r}_i +\mathbf{r}_{i+1})/2$.
The system size used is $N = 2^{24} \approx 10^{7}$, with $N_x = N_y = 2^{12}$.

We show the profile of the hopping amplitude function in \cref{fig: 8QC hopping & Bands}(a), together with a magnified view of the central region to show the eightfold patterns at the different length scales.  In \cref{fig: 8QC hopping & Bands}(b), the total spectral function along $k_x = k_y$ is shown, revealing the emergent moir\'e gaps in the electronic spectrum
created by the quasicrystalline pattern. In \cref{fig: 8QC hopping & Bands}(c), the LDOS $\rho(\mathbf{r},\omega=0)$ is shown with the same region magnified, highlighting similar patterns as the ones seen in \cref{fig: 8QC hopping & Bands}(a). 
Finally,
we show in \cref{fig: 8QC hopping & Bands}(d), the projected spectral functions $A_\text{P}(\mathbf{R},\mathbf{k},\omega)$ for the central region featuring a maximum of $t(\mathbf{r})$ and another towards the bottom right corner featuring a minmum of $t(\mathbf{r})$. These regions, of linear sizes $N_\mathbf{R}\approx N_x/16=N_y/16$, are indicated with a square in \cref{fig: 8QC hopping & Bands}(c). 
Similar to the 1D case in \cref{fig: projected spc func intqc1d}(c) and \cref{fig: projected spc func intqc1d}(d), each spatial region in 2D
features a different local electronic structure. In particular, \cref{fig: 8QC hopping & Bands}(d) shows that 
at half filling the states show a stronger localization
not at the center of the pattern, but elsewhere.
In contrast, and as observed in 
the magnified insets in \cref{fig: 8QC hopping & Bands}(d),
the
bottom of the band has a significantly higher spectral weight
in the central region. 
It is interesting to note that in 2D, nesting
conditions for the quasicrystalline pattern become more complex than in 1D, 
making the moir\'e spectral reconstructions much richer.

\subsection{Discussion}
It is worth mentioning that while the model systems considered in this work are chosen to illustrate the capabilities of the method, the framework itself is fully general and applicable to realistic material models. Any compressible tight-binding Hamiltonian\cite{Sun2025}, obtained for example from Wannierized \emph{ab initio} calculations or experimentally fitted parameters, can be encoded in the tensor-network MPO representation described in Sec.~\cref{Sec: Methods}. The subsequent KPM and QFT steps are model-agnostic, enabling the computation of momentum-resolved spectral functions for experimentally relevant twisted van der Waals heterostructures, strain-engineered moiré systems, or aperiodic super-moiré structures. In particular, the projected spectral functions $A_\text{P}(\mathbf{R},\mathbf{k},\omega)$ provide a direct theoretical counterpart to spatially resolved QTM measurements.

Finally, we note that the interactions in the examples shown simply open correlated gaps, but they are topologically trivial.
It is nevertheless worth noting that our
methodology would be capable of capturing topological states. 
Since the method operates directly on real-space tight-binding Hamiltonians encoded as MPOs, any topological phase faithfully represented at the single-particle Hamiltonian level, such as Chern insulators, can be treated with appropriate minimal modifications.
For example, one can explicitly construct a Hamiltonian featuring a domain wall between a topological and a trivial phase. In such a setup, the projected momentum-resolved spectral function would reveal the in-gap edge modes localized at the interface. Since their spectral weight is a few orders of magnitude smaller than that of bulk states, their calculation would leverage the local projectors in real-space, to focus on the regions where the edge modes are located. 
In this sense, the approach is complementary to real-space tensor-network methods for computing topological invariants~\cite{Antao2025TensorMosaics}, providing a spectroscopic perspective directly comparable to experiment. 

\begin{figure*}[!hbt]
    \centering
    \includegraphics[width=0.95\textwidth]{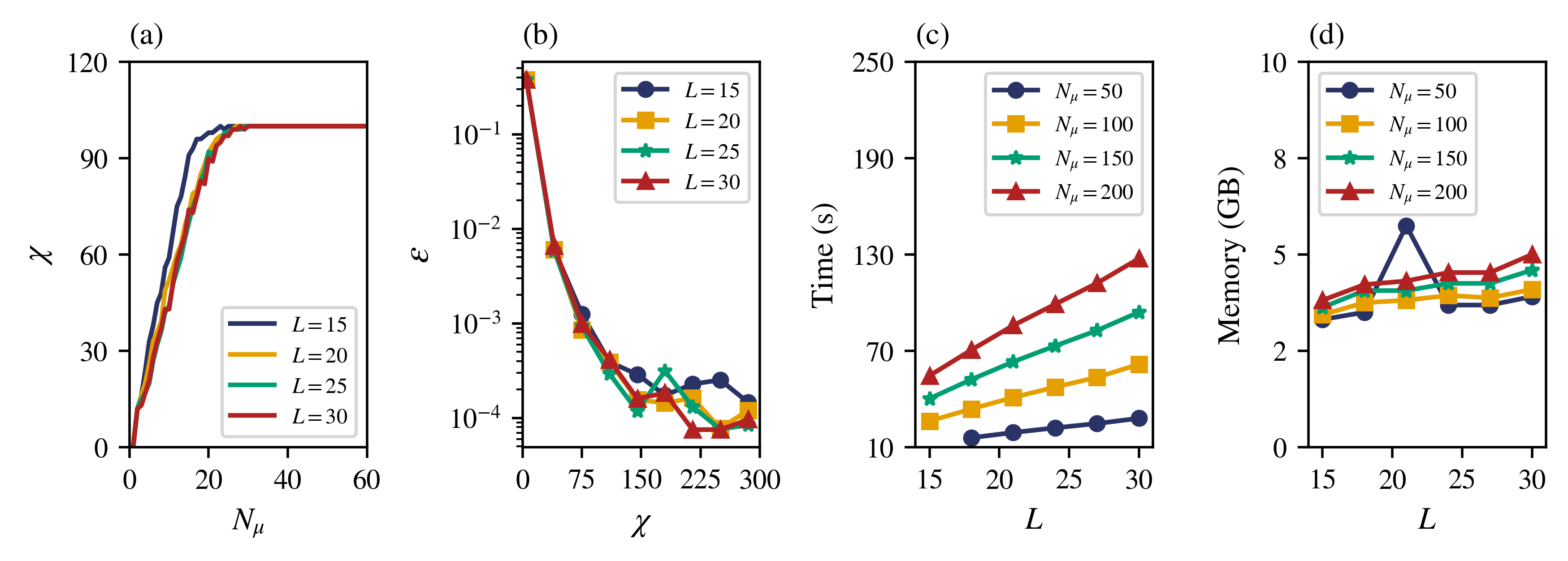}
    \caption{Benchmarking results. (a) Rank growth as a function of number of moments $N_\mu$ and logarithm of system size $L=\log_2(N)$. The maximum rank is manually set to $\chi_\text{max}=100$. (b) RMS error, defined in \cref{Eq: RMS error}, as a function of rank $\chi$ and $L$. (c) Total wall time as a function of $L$ and $N_\mu$. (d) Total memory consumption as a function $L$ and $N_\mu$.}
    \label{fig: numerical benchmarking}
\end{figure*}

\section{Numerical benchmarking}

In this appendix, we perform numerical benchmarking to provide information on the scalability of the methods developed. The most computationally expensive operation is the calculation of the Chebyshev moments of the Hamiltonian, $T_n(\hat{H})$, from \cref{Eq: delta expansion}. These involve repeated MPO--MPO contractions, whose cost grows rapidly with the maximum bond dimension and can become prohibitively expensive if left uncontrolled. Fortunately, these calculations are well suited to GPU acceleration and, when combined with systematic control of the bond dimension, result in comparatively moderate computational times. All calculations were performed on NVIDIA H200 GPUs available on the local cluster.

In \cref{fig: numerical benchmarking}, we provide a summary of the numerical benchmarking. All simulations are done for the Hamiltonian considered in \cref{Sec: One-dimensional incommensurate system under inhomogeneous strain}, with $U_\alpha=0$. The benchmarking for the SCMF calculations was performed in Ref.~\cite{Sun2025}, which is why we do not consider it here. 
We first show in \cref{fig: numerical benchmarking}(a) how the rank $\chi$ grows with the number of moments $N_\mu$ for various system sizes $N=2^{L}$, with $L=15,20,25,30$. There, one can observe that the maximum set manually to $\chi_\text{max}=100$ is reached after about $50$ moments, independent of system size. As this rank is reached, the tensor-network algorithms start to compress the MPOs substantially. 
As such, we show in \cref{fig: numerical benchmarking}(b) the error between the spectral function evaluated with a high bond dimension of $\chi_\text{max}=300$, $A(k,\omega; \chi_{\text{max}}=300)$, and $A(k,\omega; \chi_{\text{max}}=\chi)$ for arbitrary $\chi$. This is measured by the relative root-mean-square (RMS) error, defined as
\begin{equation}\label{Eq: RMS error}
    \epsilon(\chi)= \sqrt{\frac{\sum_{ij}|A_{ij}(\chi)-A_{ij}(\chi_\text{max})|^2}{\sum_{ij}|A_{ij}(\chi_\text{max})|^2}},
\end{equation}
where $A_{ij}$ abbreviates $A(k_i,\omega_j)$.
The decay of the error with increasing $\chi$ is essentially system-size independent, indicating that the accuracy of the tensor compression is governed by the chosen rank cutoff rather than by the Hilbert-space dimension. For our purposes, it is obvious that $\chi\approx100$ is more than sufficient to capture the spectral function, as the error decays subexponentially and $\epsilon(\chi\approx150)\approx10^{-4}$.  
In \cref{fig: numerical benchmarking}(c), the total wall time to calculate all the Chebyshev polynomials as a function of the logarithm of system size, $L=\log_2(N)$, is shown, for various numbers of moments $N_\mu$. The total wall-time grows linearly with L for all $N_\mu$. The slope of this linear growth increases with $N_\mu$, and, as expected, it takes more time to compute a larger number of polynomials. Finally, \cref{fig: numerical benchmarking}(d) shows the GPU memory consumption, which only increases very slightly with a higher number of moments. Additionally, the behavior appears to increase very slowly with system size. From the figure, one can estimate that the total cost of computing all the Chebyshev polynomials is between $3\,\mathrm{GB}$ and $5\,\mathrm{GB}$. 

Beyond the calculation of the moments, one must still combine them into $\delta(\omega-\hat{H})$ using \cref{Eq: delta expansion} and perform the Fourier transform. These steps take at most a few minutes. The final step is a sampling procedure, which we typically perform on a $100\times100$ grid (energy versus position in 1D, and $x$ versus $y$ at fixed energy in 2D). This stage is the most time-consuming, and scales as 
$O(N_{\mathrm{samp}})$ with the number of sampled points (here 
$N_{\mathrm{samp}} \sim 10^{4}$), and is effectively independent of the moment number 
$N_{\mu}$ and other parameters used in the moment construction. In addition, we also perform block averaging in our coarse-grained sampling procedure, which also increases the computational time. As an example, the full computational time for a noninteracting 2D system with $N=2^{24}$, without averaging, takes around two hours for $80$ $k$-points and $100$ $\omega$-points.

\section{Conclusion} \label{Sec: Conclusion}
Quantum twisting microscope experiments enable performing
momentum-resolved measurements with spatial resolution in two-dimensional materials. However, from the computational perspective,
simulating those measurements requires computing momentum-resolved
spectral functions in exceptionally large
super-moir\'e materials. 
Here, we have demonstrated a tensor network methodology to compute
momentum-resolved spectral functions in large-scale super-moir\'e systems. Our approach combining a tensor-network kernel polynomial method with
a quantum Fourier transform, 
enables the resolution of spectral features both in real and momentum space, while accommodating moir\'e and
super-moir\'e spatial modulations and electron-electron interactions treated at the self-consistent mean-field level.
We demonstrated this methodology both in one and two-dimensions,
capturing
the coexistence of interactions and non-uniform strain,
and a quasicrystalline eightfold-symmetric super-moir\'e potential.
We showed that our tensor network algorithms allows
resolving interaction-induced and
moir\'e-induced spectral gaps and the complex structure of non-periodic band
features, and enabling mapping the momentum-resolved spectral
across different regions of the system. 


Our results put forward a widely applicable methodology to compute
momentum-resolved spectral functions of exceptionally large super-moir\'e
systems.

\textbf{Acknowledgments}
We acknowledge the computational resources provided by the Aalto Science-IT project
and the financial support from InstituteQ, 
the
Research Council of Finland (project No. 370912), 
the Finnish Ministry
of Education and Culture through the Quantum Doctoral Education Pilot Program (QDOC VN/3137/2024-OKM-4), the
Finnish Quantum Flagship (project No. 358877, Aalto University),
the Finnish Centre of Excellence in Quantum Materials (No. 374166),
and the ERC Consolidator Grant ULTRATWISTROICS (Grant agreement no.
101170477). 
We thank L. Camerano, L. Eek, R. Valenti, X. Waintal,
and A. Akhmerov for useful discussions. 
The code used for this work can be consulted at \cite{anouarrepo}

\bibliography{Refs,biblio_scf,biblio_chern_marker}


\end{document}